\documentclass[aps,floatfix,prl,preprint]{revtex4-1}
\usepackage{graphicx,bm,amsmath,color,tabularx,amssymb}
\begin{document}

\def\*{^{(*)}}
\def\be{\begin{equation}}
\def\ee{\end{equation}}
\def\dq{\frac{d^3q}{(2\pi)^3}\,}
\def\brf{\mathcal B}
\def\Lc{\Lambda_c}
\def\Lp{\Lambda_{c}(2595)}
\def\S{\Sigma_c}
\def\jp{J/\psi}
\def\jpp{\jp\, p}
\def\D{\bar D}
\def\K{\bar K}
\def\>{\big>}
\def\<{\big<}
\def\|{\big\vert}
\def\rf#1{(\ref{#1})}
\def\etal{\textit{et~al.~}}
\def\MeV{\textrm{ MeV}}
\def\nb#1{{\color{red}(#1)}}
\def\blue#1{{\color{blue}#1}}
\def\etc{{\nb{\ldots}}}
\def\Pcs{P_{\psi s}^\Lambda(4338)}

\title{The LHCb state $\Pcs$ as a triangle singularity}
\author{T.\,J.\,Burns}
\affiliation{Department of Physics, Swansea University, Singleton Park, Swansea, SA2 8PP, UK.}
\author{E.\,S.\,Swanson}
\affiliation{Department of Physics and Astronomy, University of Pittsburgh, Pittsburgh, PA 15260, USA.}

\begin{abstract}
We present a model for the $J/\psi \, \Lambda$ spectrum in $B^-\to J/\psi \, \Lambda\, \bar{p}$ decays, including the $\Pcs$ baryon recently observed by the LHCb collaboration. We assume production via triangle diagrams which couple to the final state via non-perturbative interactions which are constrained by heavy-quark and $SU_3$-flavor symmetry. The bulk of the distribution is described by a triangle diagram with a color-favored electroweak vertex, while the sharp $\Pcs$ enhancement is due to the triangle singularity in another diagram featuring a $1/2^-$ baryon consistent with $\S(2800)$. We predict a comparable $\Pcs$ signal in $\eta_c \, \Lambda$, and anticipate possible large isospin mixing effects through decays to  $J/\psi \, \Sigma^0$ and $\eta_c \, \Sigma^0$.
\end{abstract}

\maketitle 

\section{Introduction}
\label{Sec:introduction}

The LHCb collaboration continues its exploration of hadronic interactions as revealed by electroweak decays of heavy hadrons, recently announcing the discovery of a signal in the $J/\psi\, \Lambda$ mass spectrum of $B^-\to J/\psi \, \Lambda\, \bar{p}$~\cite{ChenParticleZoo}. The mass and width of $\Pcs$ are \begin{align}M&= 4338.2 \pm 0.7~\MeV,\\ \Gamma &= 7.0 \pm 1.2~\MeV,\end{align}
and $J^P= 1/2^-$ quantum numbers are preferred.

Because of its proximity to $\Xi_c\D$ threshold, a molecular interpretation of $\Pcs$ has been proposed~\cite{Karliner:2022erb,Wang:2022mxy,Yan:2022wuz}. Molecules with $\Xi_c\D$ constituents have been predicted in a wide range of models, typically assuming a  binding interaction due to boson exchange, or effective field theory constrained by heavy quark symmetry \cite{Wang:2019nvm,Xiao:2019gjd,Peng:2019wys,Peng:2020hql,Chen:2020uif,Dong:2021juy,Yan:2021nio,Chen:2021cfl,Chen:2020kco,Liu:2020hcv,Zhu:2021lhd,Xiao:2021rgp,Chen:2022onm,Wang:2019nvm}. In such models, whether or a not a particular state binds is ultimately determined by one or more parameters which are fit to data, such as the form factor cut-off, or contact terms attributed to unknown short-distance physics. For this reason, the molecular approach is only robust to the extent that it can simultaneously describe several different states with the same fit parameters. With respect to this criterion, the molecular scenario for $\Pcs$ runs into problems. 

If $\Pcs$ is a $\Xi_c\D$ molecule ($1/2^-$), it could potentially have $\Xi_c\D^*$ partners ($1/2^-$ and $3/2^-$). From heavy quark symmetry, the potentials in all three channels are identical, 
\begin{align}
    V(\Xi_c\D,1/2^-)=V(\Xi_c\D^*,1/2^-)=V(\Xi_c\D^*,3/2^-),\label{eq:constraint}
\end{align}
which implies the three states should have comparable binding energies, an expectation which is confirmed in a one-boson model calculation \cite{Wang:2022mxy}. The $P_{cs}(4459)$ state observed at LHCb~\cite{Aaij:2020gdg} is a candidate for a $\Xi_c\D^*$ molecule~\cite{Peng:2020hql,Chen:2020kco,Liu:2020hcv,Zhu:2021lhd,Xiao:2021rgp}, but if it is the partner of $\Pcs$ as a $\Xi_c\D$ molecule, it implies a drastic violation of heavy-quark symmetry, since its binding energy is so large (19~MeV) compared to $\Pcs$ (which is at threshold). 

Similarly, the hypothesis that $P_{cs}(4459)$ consists of two states \cite{Aaij:2020gdg}, interpreted as $ \Xi_c\D^*$ molecules ($1/2^-$ and $3/2^-$) \cite{Karliner:2022erb,Wang:2022mxy}, is problematic, because the 13~MeV mass splitting contradicts the above expectation from heavy-quark symmetry. Although some level of splitting is expected in models (for example due to coupled-channel effects)~\cite{Wang:2019nvm,Peng:2020hql}, it remains to be seen whether such models can obtain the right level of splitting, and also reconcile the significant $\Xi_c\D^*$ binding with the lack of binding in $\Xi_c\D$. We note that in the model of ref.~\cite{Yan:2022wuz}, there is no region of parameter space in which both $\Pcs$ and two $P_{cs}(4459)$ states can be accommodated.

 Refs.~\cite{Karliner:2022erb,Wang:2022mxy} have also argued for an analogy between $\Pcs$ and $P_{cs}(4459)$ (as $\Xi_c\D\*$  molecules),  and $P_c(4312)$, $P_c(4440)$ and $P_c(4457)$ \cite{Aaij:2015tga,Aaij:2019vzc,Aaij:2016phn} (as $\S\D\*$ molecules). The analogy is misleading, considering that the corresponding potentials are neither related by $SU(3)$ flavor (as $\Xi_c$ and $\S$ belong to different flavor multiplets), nor heavy-quark spin symmetry (as  $\Xi_c$ and $\S$ have different light quark spins) \cite{Peng:2020hql}. Indeed, heavy quark symmetry implies a completely different pattern of states in $\S\D\*$ systems (where the potentials are spin-dependent~\cite{Liu:2019tjn,Liu:2019zvb,Du:2019pij,Sakai:2019qph,Valderrama:2019chc,Xiao:2019aya,Peng:2020xrf,Du:2021fmf,Liu:2020hcv}) and $\Xi_c\D\*$ systems (where they are not). Moreover, the analogy relies on the assumption that $P_c(4440)$ and $P_c(4457)$ are both $\S\D^*$ molecules, and we recently argued that this assumption is not consistent with experimental constraints \cite{Burns:2021jlu}. (Scenarios with different interpretations for $P_c(4457)$ \cite{Burns:2019iih,Burns:2022uiv} do not have the same problem.)  
 
An additional awkward feature of the molecular scenario for $\Pcs$, which has hardly been discussed in the literature, is that it is not bound with respect to the $\Xi_c\D$ thresholds, but somewhat above:
\begin{align}
\Xi_c^0\D^0&=4335.28\pm 0.33~\MeV,\\
\Xi_c^+D^-&=4337.37\pm 0.28~\MeV.
\end{align}
The situation here is similar to $P_c(4457)$, which is widely interpreted as a $\S\D^*$ bound state, despite having a mass which is consistent with the threshold not only for $\S\D^*$, but also $\Lc(2595)\D$. Signals at (rather than below) threshold are more amenable to non-resonant interpretations, and we recently demonstrated that $P_c(4457)$ can be explained as a cusp or an enhancement due to the logarithmic triangle singularity \cite{Burns:2022uiv}. In this paper we explore related possibilities for $\Pcs$.

\section{Model}

\begin{figure*}
    \centering
    \includegraphics[width=0.45\textwidth]{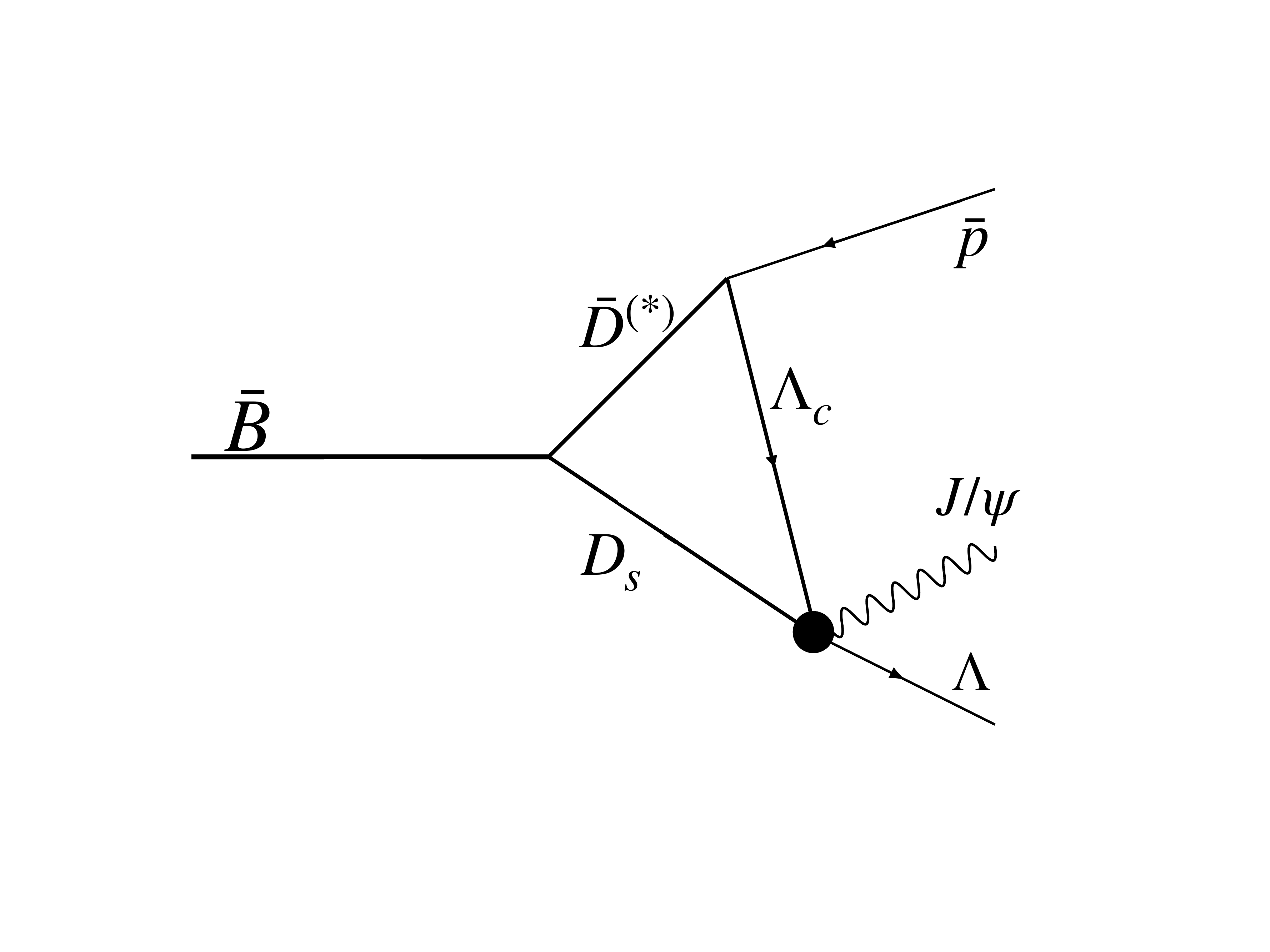}
    \includegraphics[width=.45\textwidth]{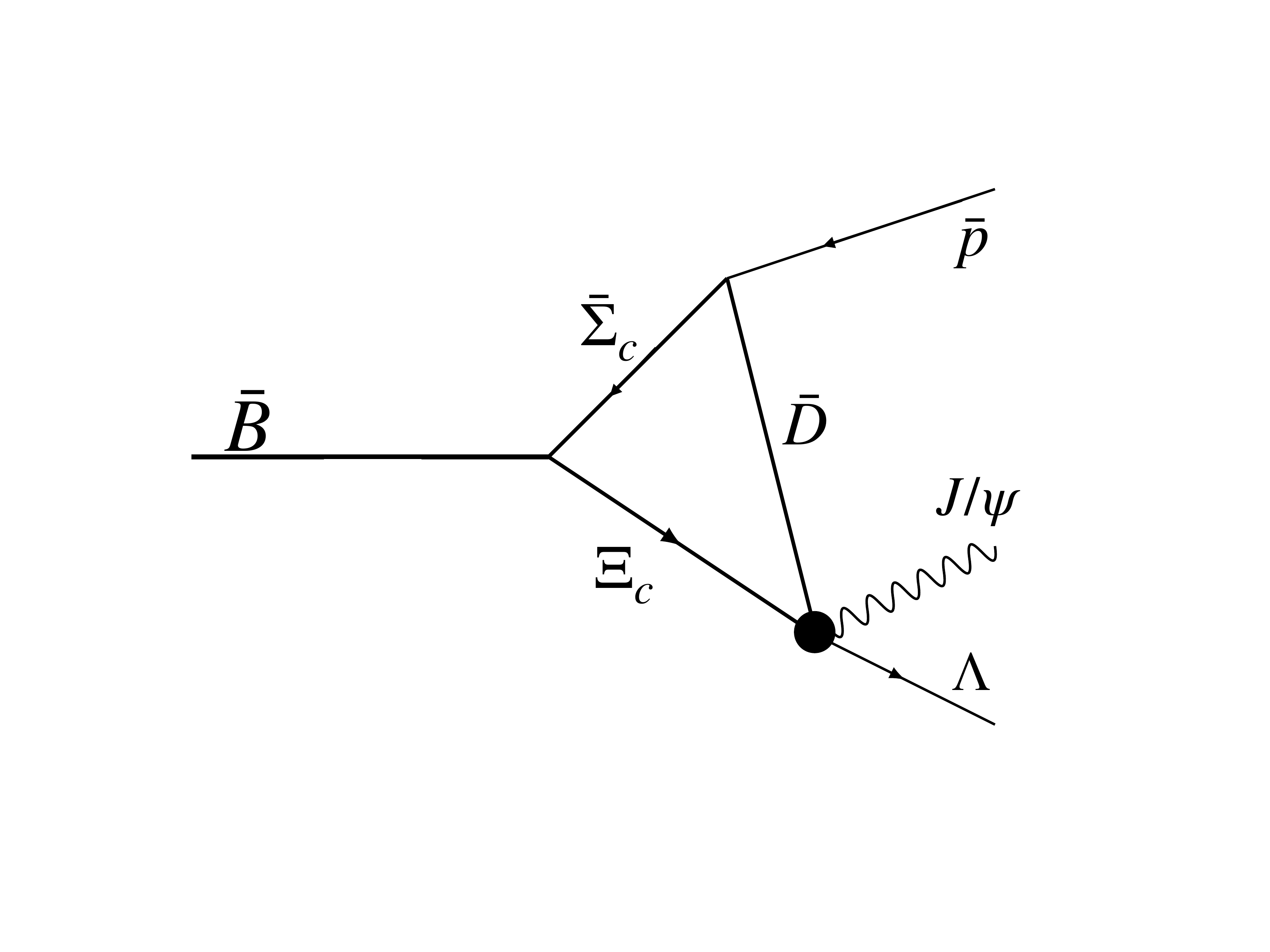}
    \caption{Diagram $T_1$ (left) has a color-favored weak transition, whereas $T_2$ (right) is color-suppressed but enhanced at the $\Xi_c\D$ threshold due to the triangle singularity. The solid circles indicate non-perturbative final-state interactions, as described in the text.}
    \label{fig:prod}
\end{figure*}

We assume, as in our previous work \cite{Burns:2020epm,Burns:2022uiv}, that the distribution can be described by triangle diagrams which couple to the final state through interactions which respect heavy-quark symmetry, and that the dominant diagrams are those with color-favored weak vertices. Hence we consider $\bar B\to D_s^{(*)-}\bar D\*$ transitions, noting that such branching fractions range from approximately 1-3\%. We can then form the triangle diagram shown in Fig.~\ref{fig:prod} (left), involving virtual $\Lambda_c$ exchange, followed by rescattering into the final $J/\psi\,\Lambda$. Generically, the distribution associated with this diagram peaks around the $\Lambda_c^+ D_s^-$ threshold (where it has a cusp). We notice that the $J/\psi \Lambda$ distribution~\cite{ChenParticleZoo} has exactly this shape, and we regard this as strong support for this proposed production mechanism. We also notice that, similar to the other exotic hadron systems~\cite{Burns:2021jlu,Burns:2022uiv}, the tree-level diagrams for the $ J/\psi \, \Lambda\, \bar{p}$ final state are color-suppressed; hence it is natural to assume that the color-favored triangle diagram is a dominant contribution.

The proximity of $\Pcs$ to $\Xi_c\D$ threshold also suggests a role for diagrams in which $\Xi_c\bar{D}$ rescatters to $ J/\psi\, \Lambda$, and here we notice an intriguing possibility. Such a novel intermediate state can be realized via an electroweak decay, such as $B^- \to \bar\Lambda_c \Xi_c$ or $B^- \to \bar\Sigma_c \Xi_c$, as shown in Fig.~\ref{fig:prod}~(right). Although the electroweak vertex is color-suppressed, we notice that the first of these modes has been observed in experiment \cite{Belle:2018kzz}, with quite significant branching fraction $(9.51\pm 2.1\pm 0.88)\times 10^{-4}$. Even accounting for some suppression at the electroweak vertex, such diagrams can make a significant contribution to the amplitude near the $\Xi\D$ threshold, if the mass of the $\bar \Lambda_c$ or $\bar \Sigma_c$ leads to a logarithmic singularity in the triangle diagram. The same idea has been applied in different contexts to explain other exotic hadron phenomena, as reviewed in ref.~\cite{Guo:2019twa}.

We will concentrate on the $\bar\Sigma_c$ (rather than $\bar \Lambda_c$) diagram, since only this is capable of producing the specific charge combination $\Xi_c^+D^-$ which, given the mass of $\Pcs$, appears to be relevant. Solving the non-relativistic version of the Landau equations \cite{Guo:2019twa}, we find that the requisite $\bar\Sigma_c$ mass is 2810~MeV. Amazingly, there is a three-star resonance $\S(2800)$ whose mass is almost exactly right. The skeptical reader may suspect
that there is a large number of possible $\S$ states, and
that we are simply choosing one with the right mass in
order to make our proposed mechanism work. On the contrary -- aside from the familiar ground states $\S(2455)$ and $\S(2520)$, the only $\S$ baryon in the Particle Data Group tables
is $\S(2800)$~\cite{ParticleDataGroup:2020ssz}.

Hence we will attempt to fit the $J/\psi\, \Lambda$ distribution, adopting the following model for the $B^-\to J/\psi \, \Lambda\, \bar{p}$ amplitude:
\begin{equation}
\mathcal{A} = b + g_1 T_1 + g_2 \frac{1}{\sqrt{6}} \left[ 2 T_2^{(--)} - T_2^{(-)}\right],
\label{eq:A}
\end{equation}
where $b$ (the background, a complex constant) and $g_{1,2}$ (the production couplings) are fit to data, and $T_1$ and $T_2$ are the  sub-amplitudes corresponding to the diagrams in the left and right panels of Fig.~\ref{fig:prod}, respectively, computed as described below.

Because the $\Pcs$ peak is much closer to the threshold for $\Xi_c^+D^-$ rather than $\Xi_c^0\D^0$, we do not expect isospin symmetry to be respected in this system. This was discussed in refs~\cite{Yan:2022wuz,Meng:2022wgl}, and is similar to related observations in, for example, $X(3872)$ and  the $P_c$ states~\cite{Swanson:2003tb,Close:2003sg,Burns:2015dwa,Guo:2019fdo,Guo:2019kdc}. In $T_2$ we therefore consider the $\Xi_c^+D^-$ and $\Xi_c^0\D^0$ diagrams separately, and refer to these as $T_2^{(--)}$ and $T_2^{(-)}$, corresponding to $\bar\Sigma_c^{--}$ and $\bar\Sigma_c^-$ states in the triangle diagrams. The remaining factors in equation~\eqref{eq:A} arise from an isospin decomposition of the amplitude. Notice that the weighted combination of $T_2^{(--)}$ and $T_2^{(-)}$ corresponds to $\Xi_c\D$ in a linear combination of isospin 0 and 1. Hence the production mechanism itself does not respect isospin, even before considering the mass difference between $\Xi_c^+D^-$ and $\Xi_c^0\D^0$. This is a different (and more significant) source of isospin mixing than that due to the $\Xi_c\D$ masses; similar effects could be present in the $X(2900)$ system~\cite{Burns:2020xne}.

Our calculation of the amplitudes $T_1$ and $T_2$ follows the method outlined in detail in our previous paper~\cite{Burns:2022uiv}, so here we just mention some key points. The triangle diagrams are computed in the nonrelativistic limit, incorporating form factors for the strong decay vertex and the final state interactions. The form factor scale is fixed at 800~MeV, as in the previous study. Because modeling  the electroweak vertex is difficult, and because it does not vary much over the rather narrow  phase space available, we have chosen to employ a constant electroweak vertex, whose strength is ultimately absorbed into production couplings $g_1$ and $g_2$ which, as discussed below, are fit to data. 

In diagram $T_1$, the functional dependence of the amplitude is insensitive to the choice of the ``$\D\*$'' mass. Hence instead of separately considering contributions from $\D$ and $\D^*$, we compute a single diagram, using the physical $\D^*$ mass.

In diagram $T_2$, we notice that, in order to generate a prominent enhancement at $\Xi_c\D$ threshold, we need the $\bar\Sigma_c\to\D\bar p$ vertex to be S-wave. The corresponding  $\S$ state therefore has $J^P$ quantum numbers $1/2^-$. The quantum numbers of $\S(2800)$ have not been measured, but the mass is consistent with expectations for the 1P multiplet, which includes states with quantum numbers $1/2^-$, $3/2^-$ and $5/2^-$ \cite{Copley:1979wj,Chen:2007xf,Cheng:2006dk,Cheng:2015naa,Wang:2021bmz}. We therefore assume that there is a $1/2^-$ state around 2800~MeV, though we are not necessarily assuming it is $\S(2800)$ itself. For the purposes of the calculation, we fix the ``$\bar\Sigma_c$'' mass to 2801~MeV (the measured mass of $\S(2800)^{++}$~\cite{ParticleDataGroup:2020ssz}), but we notice that the fit quality is not highly sensitive to this choice. (Although the logarithmic singularity strictly arises within a specific and narrow window of $\bar\Sigma_c$ masses, in practice we find very strong enhancements in the triangle diagram over a range of $\bar\Sigma_c$ masses around 2800~MeV.)

Most particles in the triangle diagrams are very narrow, so we ignore their widths and instead introduce a small imaginary part $\epsilon$ in the energy denominators. The exception is $\bar\Sigma_c$, whose width is included explicitly. We consider two cases: $\Gamma=70$~MeV (from the measured width of the $\S(2800)$), and $\Gamma=15$~MeV (a model prediction for the width of the $1/2^-$ state in the 1P multiplet~\cite{Wang:2021bmz}).

The triangle diagrams couple to the $J/\psi\, \Lambda$ final state via non-perturbative final-state interactions, represented by the solid circles in Fig.~\ref{fig:prod}. Following ref.~\cite{Burns:2022uiv}, we assume a separable form for the interaction potential, and so obtain the non-perturbative T-matrix by solving the Bethe-Heitler equation, $T= V + VGT$, using algebraic methods. The final-state interactions are responsible for couplings among all the relevant channels so far discussed ($\Lc^+D_s^-$, $\Xi_c^+D^-$, $\Xi_c^0 \D^0$, $\Lambda J/\psi$), as well as others which are related to these by heavy-quark symmetry and which also couple to  $1/2^-$ in  S-wave. In particular, we include $\Lambda \eta_c$, $\Sigma J/\psi$ and $\Sigma\eta_c$ as possible final states of interest, noting that the $\Sigma$ modes are relevant because of the explicit isospin mixing in the model. We do not include other channels such as $\Lc D_s^*$, $\Xi_c\D^*$ and $\Xi_c^{(\prime,*)}\D\*$, whose thresholds are beyond the kinematic boundary for $J/\psi \Lambda$ in $B^-\to J/\psi \, \Lambda\, \bar{p}$.

We  choose to model the final-state interactions as contact terms constrained by heavy-quark symmetry, following an approach which has been widely applied to $\S\*\D\*$ systems~\cite{Liu:2018zzu,Liu:2019tjn,Sakai:2019qph,Valderrama:2019chc,Liu:2019zvb,Du:2019pij,Peng:2020xrf,Du:2021fmf}, and more recently $\Xi_c^{(\prime,*)}\D\*$ systems \cite{Peng:2019wys,Yan:2021nio,Peng:2020hql,Liu:2020hcv,Yan:2022wuz}. We previously tabulated the relevant contact terms for S-wave interactions among isospin 1/2 channels $\Lc\D\*$, $\S\*\D\*$, $N\jp$ and $N\eta_c$~\cite{Burns:2022uiv}. Most of the contact terms we need for the present case can be extracted from those by assuming, as in refs~\cite{Peng:2020hql,Yan:2021nio,Yan:2022wuz}, that the interactions are invariant under rotations in $SU_3$ flavor space. Hence the matrix elements for the $\|SU_3\textrm{~flavor},SU_2\textrm{~isospin}\>$ states $\|\mathbf 8,\mathbf 1\>$ and $\|\mathbf 8,\mathbf 3\>$, formed out of $\Lc^+D_s^-$, $\Xi_c^+D^-$, $\Xi_c^0 \D^0$ using $SU_3$ isoscalar factors~\cite{Kaeding:1995vq}, are identical to those of $\Lc\D$, which is the $\|\mathbf 8, \mathbf 2\>$ state with the same spin structure. In a similar way, matrix elements involving $\Lambda\jp$ and $\Sigma\jp$ ($\Lambda\eta_c$ and $\Sigma\eta_c$) are the same as the corresponding terms in our previous paper involving $N\jp$ ($N\eta_c$). In summary, we have
\begin{align}
    \<\mathbf 8,\mathbf1\|V\|\mathbf 8,\mathbf 1\>&=\<\mathbf 8,\mathbf 3\|V\|\mathbf 8,\mathbf 3\>=A,\\
    \<\mathbf 8,\mathbf1\|V\|\Lambda J/\psi\>&=\<\mathbf 8,\mathbf 3\|V\|\Sigma J/\psi\>=\frac{\sqrt 3}{2}D,\\
     \<\mathbf 8,\mathbf1\|V\|\Lambda \eta_c\>&=\<\mathbf 8,\mathbf 3\|V\|\Sigma \eta_c\>=\frac{1}{2}D,
\end{align}
where $A$ and $D$ are contact terms which are (somewhat) constrained by our previous analysis of $\Lambda_b\to \jpp K^-$ decays \cite{Burns:2022uiv} (see below). 

The $\Lc^+D_s^-$, $\Xi_c^+D^-$, $\Xi_c^0 \D^0$ basis states also combine into a flavor singlet, which implies an additional contact term which is independent of the other contact terms \cite{Yan:2021nio}. We call this $A'$, by analogy with $A$ above:
\begin{align}
    \<\mathbf 1,\mathbf 1\|V\|\mathbf 1,\mathbf 1\>=A'.
\end{align}

Although the potentials are assumed to respect $SU_3$ and $SU_2$ symmetries, the transition amplitudes will not, because of mass differences among the constituents with different flavors. Hence we formulate the potential in the particle basis, and show the result in Table~\ref{tab:V}, where we have introduced $\Delta = (A'-A)/3$. Note that, as in our previous work, we are assuming that potentials coupling two ``closed-charm'' states (such as $\Lambda J/\psi\to\Lambda J/\psi$) are zero.
\begin{table}
\begin{tabularx}{\columnwidth}{X|XXX|XXXX}
\hline
 & $\Lambda_c^+ D_s^-$ & $\Xi_c^+ D^-$  & $\Xi_c^0 \bar{D}^0$ & $\Lambda  J/\psi$  & $\Lambda \eta_c$ & $\Sigma J/\psi$ & $\Sigma \eta_c$ \\
\hline
$\Lambda_c^+ D_s^-$   &  $A+\Delta$  &  $\Delta$   &   $-\Delta$  &   $\frac{D}{\sqrt{2}}$  & $\frac{D}{\sqrt{6}}$ & 0 & 0 \\
$\Xi_c^+ \bar{D}^-$   &     &  $A+\Delta$   &   $-\Delta$   &  $-\frac{D}{2 \sqrt{2}}$ & $-\frac{D}{2 \sqrt{6}}$ & $\frac{\sqrt{3} D}{2 \sqrt{2}}$ & $\frac{D}{2 \sqrt{2}}$ \\ 
$\Xi_c^0 \bar{D}^0$  &   &     &  $A+\Delta$   &   $\frac{D}{2 \sqrt{2}}$ & $\frac{D}{2 \sqrt{6}}$ & $\frac{\sqrt{3} D}{2 \sqrt{2}}$ & $\frac{D}{2 \sqrt{2}}$ \\ 
\hline
$\Lambda  J/\psi$     &     &      &    &  0 & 0 & 0 & 0 \\
$\Lambda  \eta_c$   &       &      &       &  & 0 & 0 & 0   \\
$\Sigma J/\psi$   &       &      &       &   & & 0 & 0   \\
$\Sigma  \eta_c$   &       &      &      & & &  &  0   \\
\hline
\end{tabularx}
\caption{Contact terms in the $1/2^-$ channel.}
\label{tab:V}
\end{table}

With our model for the potentials, analysis of  $\Lambda_b\to \jpp K^-$ decays \cite{Burns:2022uiv} indicates that $D$ is constrained very roughly to be a number of order 1 GeV$^{-2}$, and we will adopt that value in this work. The contact term $A$ is not well-constrained by $\Lambda_b\to \jpp K^-$ decays, although we know that it cannot be large and negative, as it would imply $\Lc\D\*$ bound states, which are apparently not seen in the data. Hence in this work we consider $A\ge 0$. 

\section{Results}
We first attempt a fit with only diagram $T_1$ (fixing $g_2=0$), corresponding to the conventional molecular scenario. To understand the possibilities, we note that in the isospin basis, the diagonal $\Xi_c\D$ potentials are 
\begin{align}
    V(\Xi_c\D,I=0)&=A+2\Delta,\label{eq:1}\\
    V(\Xi_c\D,I=1)&=A.\label{eq:2}
\end{align}
Comparing to the $\Lc^+D_s^-$ potential in Table~\ref{tab:V}, it suggests there may be some region of parameter space (with $\Delta<0$) in which $\Xi_c\D$ ($I=0$) binds, but not $\Lc^+D_s^-$. But in practice we find the opposite, namely a prominent signal at (or below) $\Lc^+D_s^-$ threshold, rather than $\Xi_c\D$. One reason is that $\Delta$ not only contributes to the diagonal potential, but also the off-diagonal coupling between  the $\Lc^+D_s^-$ and $\Xi_c\D$ channels,
generating an effective attraction in the lower channel ($\Lc^+D_s^-$), and also broadening any peak associated with $\Xi_c\D$ (increasing its decay width). Ref~\cite{Yan:2022wuz} also observed that binding in $\Xi_c\D$ inevitably also implies binding in $\Lc^+D_s^-$. We also remark that, in our model, the $\Lc^+D_s^-$ state is always more prominent in the amplitude, since it is produced directly in the triangle diagrams -- in contrast to $\Xi_c\D$, which arises in diagram $T_1$ only through final-state interactions (via $\Delta$). All of this is problematic for the molecular scenario.

\begin{figure}[ht]
    \centering
    \includegraphics[width=\textwidth]{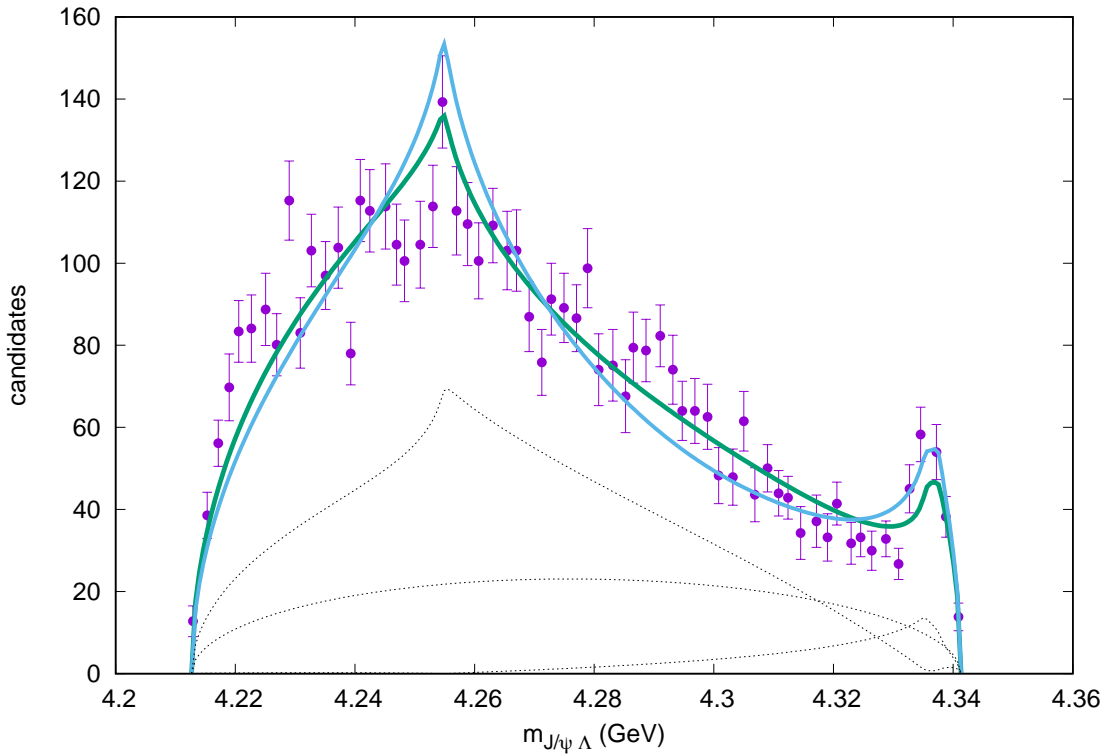}
    \caption{The $J/\psi \Lambda$  invariant mass spectrum for parameter sets A (green) and B (blue), compared to the experimental data from ref.~\cite{ChenParticleZoo}. The thin dotted lines show the separate contributions of $T_1$, background,  and $T_2$  (top to bottom), for parameter set A.
     }
    \label{fig:fit1}
\end{figure}

Hence we proceed with the full model, including both diagrams $T_1$ and $T_2$. The results are much better: in Fig.~\ref{fig:fit1} we give two illustrative examples of successful fits where, in both cases, the bulk of the distribution is captured by diagram $T_1$, while the $\Pcs$ peak is from diagram $T_2$, due to the anticipated triangle singularity. The shape of the triangle peak is sensitive to the chosen width of the virtual $\bar\Sigma_c$, which in turn has implications for the contact terms $A$ and $\Delta$. The two parameter sets illustrated in Fig.~\ref{fig:fit1} are summarized in Table~\ref{tab:par}.

\begin{table}
    \centering
    \begin{tabularx}{\columnwidth}{XXXl}
    \hline
         &  $\Gamma(\bar\Sigma_c)$  / MeV&$A$ / GeV$^{-2}$ &$\Delta$ / GeV$^{-2}$\\
         \hline
Set A   & 70&6&$-7$\\
Set B   & 15& 0&$-1$\\
\hline
    \end{tabularx}
    \caption{Parameter sets A (green) and B (blue) in Fig~\ref{fig:fit1}.}
    \label{tab:par}
\end{table}

The effect of a broader $\bar\Sigma_c$ is to broaden the triangle peak in diagram $T_2$. Hence in Set~A, in order to obtain a sufficiently sharp peak, we adjust $A$ and $\Delta$ to effectively add some attraction to the $\Xi_c\D$ channel, but in such a way that we do not have a bound state in $\Xi_c\D$ or $\Lc^+D_s^-$. With the narrower $\bar\Sigma_c$ of Set~B, the triangle peak is already sharp enough that no extra attraction is really needed, although with our chosen $\Delta = -1$~GeV$^{-2}$ there is a small attraction, as well as coupling between the $\Xi_c\D$ and $\Lc^+D_s^-$ channels. Note that the values of $A$ and $\Delta$ are not very well constrained by the fits.



To better understand our results, let us consider one of the fits (Set A) in some more detail. The fit parameters are
\begin{align}
b& = 0.93 + 6.63 i \ \textrm{GeV}^{-2},\\
g_1& =  5766, \\
g_2& =  -316.1.
\end{align}
In Fig.~\ref{fig:fit1} we show separately (with thin dotted lines) the contributions from diagrams $T_1$ and $T_2$, and the background. Notice that $T_1$ nicely captures the overall shape of the distribution, including the peak around the $\Lc^+D_s^-$ threshold which, as mentioned previously, is a natural consequence of the assumed dominance of the color-favored triangle diagram. While it is amusing that the peak corresponds to the single data point near 140 candidates, we regard this as a fluke. It is also reassuring that the background is relatively small compared to $T_1$, and this is as expected, since tree-level production of the final state is color-suppressed.

We are also satisfied that the fit has $|g_2|<<|g_1|$, since it is consistent with expectations that the electroweak vertex in $T_2$ is suppressed compared to $T_1$ by at least the number of colors, with further suppression expected due to the orbital excitation in $\bar\Sigma_c$. As has been anticipated, despite a small coupling $g_2$, diagram $T_2$ still makes a prominent contribution to the fit because of the enhancement due to the triangle singularity.

The chi-squared value for the fit with Set A parameters is 1.9. This can be improved substantially by using a more complicated background model; however, our purpose is not to obtain a very precise fit, but to establish the physical mechanism that explains the data. 

From the matrix elements in Table~\ref{tab:V}, we can make some general observations about the prospects of observing $\Pcs$ in other final states. (Detailed predictions for distributions in various final states are difficult, without a model for the backgrounds in each case.)

For example, in $B^-\to \eta_c\, \Lambda\, \bar{p}$ decays, we expect a $\Pcs$ signal in the $\eta_c\, \Lambda$ distribution, but suppressed in comparison to the $J/\psi\,\Lambda$ signal by a factor of approximately 3 (in rate).

As noted previously, we expect $\Pcs$ to exhibit isospin mixing, as the $\Xi_c\D$ pair in diagram $T_2$ is a linear combination of isospin 0 and 1. If we ignore the (small) additional contribution to mixing due to the charged/neutral mass difference, and assume that $\Pcs$ is solely due to the triangle singularity (in the sense that there are no non-perturbative final state interactions in $\Xi_c\D$), we find that in diagram $T_2$, the signals in isospin 1 modes ($J/\psi\, \Sigma^0$ and $\eta_c \Sigma^0$) are suppressed compared to the corresponding isospin 0 modes ($J/\psi\, \Lambda$ and $\eta_c \Lambda$), but only by a factor of 3 in rate (before accounting for phase space differences). Isospin mixing in this system is therefore a very significant effect. Moreover, the $\Pcs$ peak will feature more prominently in the $J/\psi\, \Sigma^0$ and $\eta_c \Sigma^0$ distributions since (at leading order) they receive no contribution from the diagram $T_1$, which dominates the $\jp\Lambda$ spectrum.

But the factor of three only applies in the perturbative limit, and indeed we find that non-perturbative final state interactions can change the outcome considerably. For example, in parameter set A (Table~\ref{tab:par}), the values of $A$ and $\Delta$ imply attraction in isospin 0, but repulsion in isospin 1 -- see equations~\eqref{eq:1} and \eqref{eq:2}. So by introducing final state interactions to enhance the triangle peak in isospin 0, we effectively suppress the peak in isospin 1. Indeed we have verified that with our parameter set A, the suppression of the isospin 1 peak is very much stronger than the factor of 3 which applies in the perturbative limit. 

Because the magnitude of isospin mixing is correlated with the parameters $A$ and $\Delta$, future experimental measurement of isospin 1 modes could be used to constrain these parameters, which cannot be determined with current data.

\section{Conclusion}
We have demonstrated that a simple model based on the assumed dominance of color-favored weak transitions, and approximate heavy-quark and $SU_3$ flavor symmetries, can describe the  $B^-\to J/\psi \, \Lambda\, \bar{p}$ data, including the prominent $\Pcs$ peak. Notably, we are not assuming a molecular nature for $\Pcs$ -- on the contrary, we showed that the molecular scenario does not work. Our  conclusions (and methods) are very similar to those of our recent analysis of  $\Lambda_b\to\jpp K^-$ data \cite{Burns:2021jlu,Burns:2022uiv}, in which we argued that the $\S\D^*$ molecular interpretation of $P_c(4457)$ is problematic, but that several viable alternatives give an excellent fit to data. Our results underline the importance of exploring alternatives to the prevailing molecular interpretation of states which are located at thresholds.



\acknowledgments
Swanson's research was supported by the U.S. Department of Energy under contract DE-SC0019232.

\bibliography{biblio}

\end{document}